\documentclass[a4paper,12pt]{article}
\pdfoutput=1

\usepackage{amsfonts,amsmath,amssymb,bm,eucal}
\usepackage[english]{babel}
\usepackage[T1]{fontenc}
\usepackage{ae,aecompl}

\usepackage{graphicx,graphics,color,psfrag,subfigure,epsfig,epstopdf}
\usepackage{hyperref}

\usepackage[comma,numbers,sort&compress]{natbib}

\title{
\vspace*{-2cm}
\begin{flushright}
\normalsize{BARI-TH/2012-649}
\end{flushright}
\vspace*{1.5cm}
\textbf{In-medium hadronic spectral functions through the soft-wall holographic model of QCD}\\ ~\\
\author{P.~Colangelo$^a$, F.~Giannuzzi$^{a,b}$, S.~Nicotri$^{a,b}$\\
~\\ 
\normalsize\emph{$^a$Istituto Nazionale di Fisica Nucleare, Sezione di Bari, Italy}\\
\normalsize\emph{$^b$Dipartimento di Fisica, Universit\`a degli Studi di Bari, Italy}
}}

\date{}

\begin{document}

\maketitle

\begin{abstract}
We study the scalar glueball and vector meson spectral functions in a hot and dense medium by means of the soft-wall holographic model of QCD. 
Finite temperature and density effects are implemented through the AdS/RN metric.
We analyse the behaviour of the hadron masses and widths in the $(T,\mu)$ plane, and compare our results with the experimental ones and with other theoretical determinations.
\end{abstract}

\section{Introduction}

The properties of hadrons in a thermal and dense medium are currently widely investigated, both experimentally and theoretically. 
In this respect, the heavy-ion experiments at RHIC and LHC are producing a huge amount of data. 
However, from the theoretical point of view, it is still difficult to analyse these issues since one has to face nonperturbative calculations. 

The study of the spectral functions deserves particular attention since important information can be inferred from them, namely the behaviour of hadron masses and widths with temperature and density, or the highest temperature or density at which a state can exist. 
Moreover, from the vector meson spectral functions at low energy it is possible to compute transport coefficients using Kubo relations, together with other hydrodynamical quantities which can help in characterising the medium of quarks and gluons and in interpreting the experimental results \cite{Teaney:2001av}.

Hadronic spectral functions at finite quark chemical potential have been investigated by the Nambu-Jona Lasinio model \cite{Hansen:2006ee,Muller:2010am}, QCD sum rules \cite{Kwon:2008vq,Hatsuda:2008is} and effective models \cite{Urban:1999im,Post:2003hu,Santini:2008pk,Riek:2010gz}. 
On the other hand, this is a hard problem to deal with using lattice simulations, because of the sign problem occurring at nonzero chemical potential. 
Some analyses of vector mesons and scalar glueballs have been carried out by lattice simulations with two colours, where the sign problem is absent \cite{Muroya:2002ry,Hands:2007uc,Lombardo:2008vc}. 
A study with three colours in a mean field approximation at strong coupling has appeared in \cite{Kawamoto:2007cc}. 
There are also experimental evidences of the modifications of $\rho$ spectral function in medium \cite{Ozawa:2000iw,Huber:2003pu,Naruki:2005kd,Arnaldi:2006jq,Adamova:2006nu,Wood:2008ee}, as we discuss in Sect.~\ref{sec:vec}; reviews on theoretical and experimental results can be found in \cite{Rapp:2010sj,Rapp:1999ej}. 

To address the nonperturbative problem of computing in-medium hadronic spectral functions, we use the AdS/QCD soft-wall model \cite{sw}, which has been recently developed following the holographic bottom-up approaches to QCD inspired by the AdS/CFT (or gauge/gravity) correspondence \cite{maldacena1,Witten:1998qj,gubser-klebanov-polyakov}. 
In these phenomenological approaches, gauge-invariant QCD operators are related to specific five-dimensional fields living in an anti-de Sitter (AdS) space, under the assumption that the strongly coupled gauge theory defined in a Minkowski space is dual to a semiclassical gravity theory in an AdS space. 
The duality consists in the equality of the partition functions of the two theories, under suitable boundary conditions, so that correlation functions of the strongly coupled theory can be obtained deriving a semiclassical functional \cite{Witten:1998qj}.
This relation between a perturbative and a nonperturbative theory makes the correspondence interesting, in particular in view of the application to QCD.

A key feature, that any holographic model aiming at describing QCD must incorporate, is the inclusion of a mass scale that breaks conformal invariance and is related to the scale of confinement.
In the soft-wall model this is achieved by inserting in the action a dilaton background field containing a mass parameter \cite{sw}. 
Other studies of hadronic spectral functions at finite chemical potential in so-called top-down holographic models can be found in \cite{Erdmenger:2007ja,Mas:2008jz}.
We shall focus on scalar glueballs and light vector mesons, but the method we adopt is quite general and can be applied to other states.

\section{Scalar glueball spectral function}

At $T=0$ and $\mu=0$ scalar glueballs can be described in the soft-wall model by introducing a massless scalar field $X(x,z)$, dual to the QCD operator $\beta(\alpha_s)G_{\mu\nu}^aG^{a\,\mu\nu}$ \cite{Colangelo:2007pt,Forkel:2007ru}, in the action:
\begin{equation}\label{action_gl}
S=\frac{1}{2\, k^\prime} \int d^4x \,dz \,\, e^{-\phi(z)}\sqrt{g}\,g^{MN}\left(\partial_M X\right)\left(\partial_N X\right)\,.
\end{equation}
$\phi(z)$ is the dilaton term of the soft-wall model, and $g$ is the determinant of the metric of the AdS$_5$ space
\begin{equation}
ds^2=\frac{R^2}{z^2}\left(dt^2-d\bar x^2-dz^2\right)\,,
\end{equation} 
with $R$ the AdS$_5$ radius; in the following we set $R=1$. The holographic coordinate $z$ runs in the range $\epsilon \leq z < +\infty$, with $\epsilon \to 0^+$.
The dilaton profile is $\phi(z)=c^2z^2$, and the mass parameter $c$ is fixed from the $\rho$ meson mass predicted in the model: $c=m_\rho/2=0.388$ GeV. 
The quadratic form of the profile has been chosen as the simplest one allowing to reproduce Regge behaviour for light hadron masses, and to get analytical results in several sectors, mostly at $T=0$.
The parameter $k^\prime$ is fixed from the matching with QCD of two-point function \cite{Colangelo:2007if}: $k^\prime=\pi^4/(16\alpha_s^2\beta_1^2)$ (with $\beta_1$ given in QCD by 
$\beta_1=-11N_c/6+N_f/3$).

In order to study the scalar glueball spectral function at finite temperature and density, we follow the approach of \cite{Colangelo:2009ra,Fujita:2009wc,Miranda:2009uw}, where the case of finite temperature and zero density have been considered; the results obtained and the tools developed can be generalised to the finite density case in a straightforward way.

When we switch temperature and quark chemical potential on, the action for the field $X(x,z)$ is still \eqref{action_gl}, but with a different metric, which gathers information about temperature and density through a black-hole \cite{Colangelo:2010pe}:
\begin{equation}
 ds^2=\frac{R^2}{z^2}\left(f(z)dt^2-d\bar x^2-\frac{dz^2}{f(z)}\right) \qquad\qquad 0<z<z_h
\end{equation}
\begin{equation}
 f(z)=1-\left( \frac{1}{z_h^4} +q^2\, z_h^2 \right)z^4+q^2 z^6
\end{equation}
where $z_h$ is the position of the outer horizon of the black-hole, i.e. the lowest value which satisfies $f(z_h)=0$. 
This is known as AdS/Reissner-Nordstr\"om metric (AdS/RN).
The temperature of the black-hole is
\begin{equation}
 T=\frac{1}{4\pi}\left| \frac{df}{dz}\right|_{z=z_h}=\frac{1}{\pi z_h} \big(1-\frac{Q^2}{2} \big)
\end{equation}
being $Q=q z_h^3$, with $0\leqslant Q\leqslant \sqrt{2}$. 
The parameter $q$ represents the charge of the black-hole, and is related to the chemical potential $\mu$. 
Let us clarify briefly this point.
In the QCD generating functional, the quark chemical potential $\mu$ is the coefficient multiplying the quark number operator $q^\dagger q=\bar q\gamma^0q$. 
In the dual theory, such a coefficient is the source of the field associated to the quark number operator, namely the time component of a $U(1)$ gauge field $A_M$. 
The AdS/RN metric is the result of the interaction of this field with the AdS space. 
Instead of solving coupled differential equations involving the metric, the dilaton and the gauge field, as in \cite{Colangelo:2010pe,Lee:2009bya} we use the ansatz:
\begin{equation}
 A_0(z)=\mu-\kappa\, q\, z^2 \qquad\qquad A_i(z)=0,~~i=1,2,3,z\,,
\end{equation}
and, from the condition $A_0(z_h)=0$, we find
\begin{equation}
\mu=\kappa Q/z_h\,.
\end{equation}
$\kappa$ is a dimensionless parameter of the model, which must scale as $\sqrt{N_c}$ and can be determined studying different observables \cite{Lee:2009bya}; in the following we use $\kappa=1$.
For convenience, we set $c=1$, so all the following results are given in units of $c$.

\begin{figure}[h!]
 \centering
\subfigure{\includegraphics[width=6.7cm]{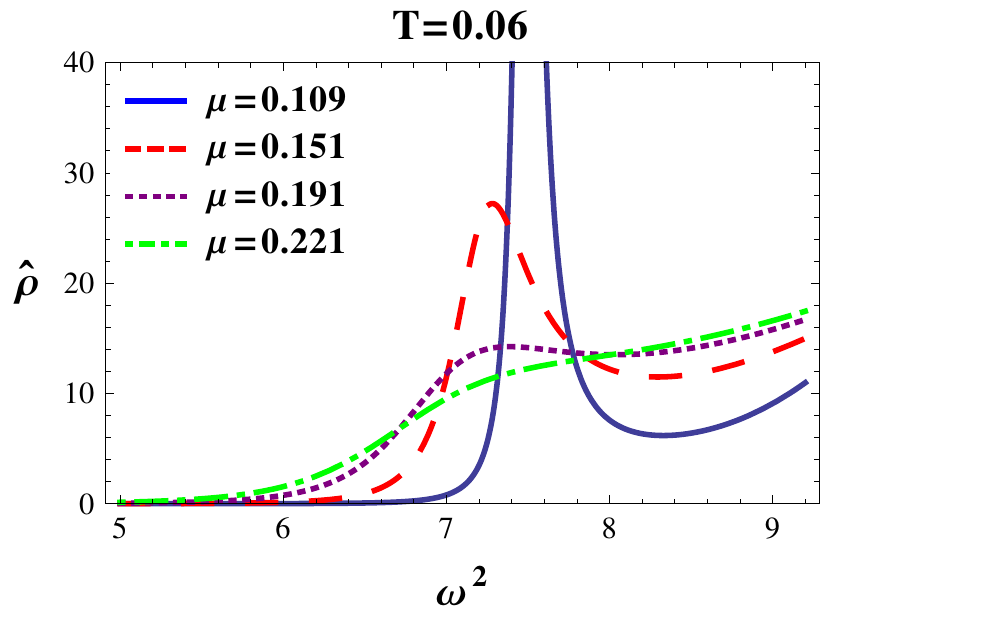}}
\subfigure{\includegraphics[width=6.7cm]{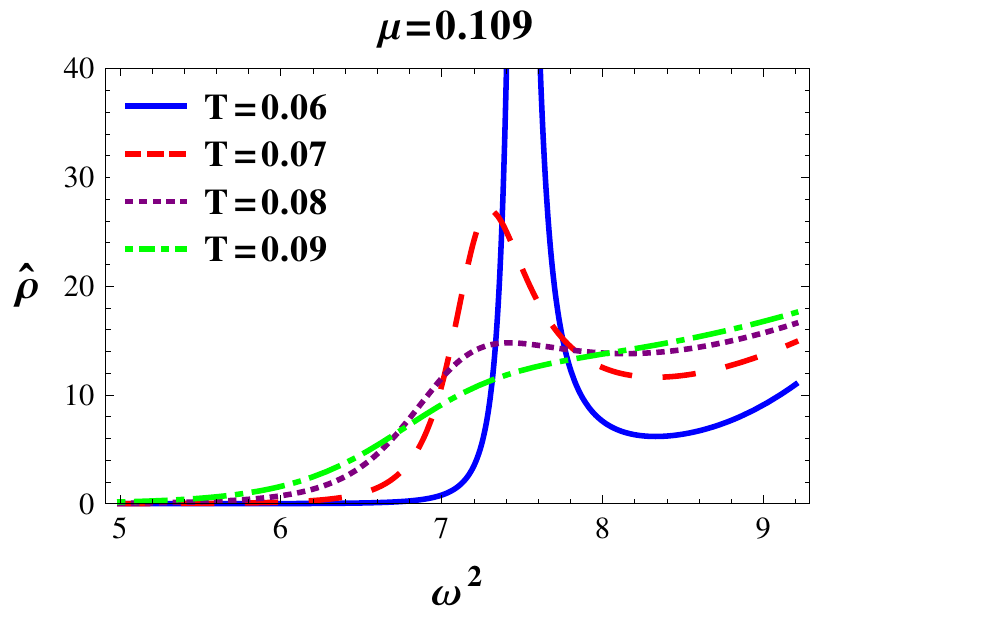}}
 \caption{Left: first peak of the scalar glueball spectral function at $T=0.06$ and $\mu=0.109$ (plain, blue), $\mu=0.151$ (dashed, red), $\mu=0.191$ (dotted, purple), $\mu=0.221$ (dash-dotted, green). Right: first peak of the spectral functions of the scalar glueball at $\mu=0.109$ and $T=0.06$ (plain, blue), $T=0.07$ (dashed, red), $T=0.08$ (dotted, purple), $T=0.09$ (dash-dotted, green). The variable $\hat\rho$ is defined as $\hat\rho=k^\prime\, \rho$. All quantities are in units of $c$.}
 \label{fig:SFglue}
\end{figure}

The bulk-to-boundary propagator $\tilde K(p,z)$ of the glueball field in $4D$ Fourier space is defined by $\tilde{X}(p,z)=\tilde K(p,z) \tilde X_0(p)$, being $\tilde X(p,z)$ and $\tilde X_0(p)$ the bulk field and the source of the QCD operator, respectively. 
From the action \eqref{action_gl} the equation of motion for $\tilde K(p,z)$ can be computed 
\begin{equation}\label{eomglue}
 \tilde K''(p,z)- \frac{4-f(z)+2c^2z^2f(z)}{z f(z)} \tilde K'(p,z)+\left( \frac{p_0^2}{f(z)^2}-\frac{\bar{p}^2}{f(z)} \right) \tilde K(p,z)=0 \,;
\end{equation}
as in \cite{Colangelo:2009ra}, the case $\bar p=0$ will be considered, so $\omega^2=p_0^2$.
The solution must satisfy two boundary conditions: $\tilde K(\omega^2,0)=1$ and
\begin{equation}
 \tilde K(\omega^2,z) \xrightarrow[]{z\to z_h} (1-z/z_h)^{-i\frac{\sqrt{\omega^2}\, z_h}{2\, (2-Q^2)}} \left(1+{\CMcal O}(1-z/z_h)\right)\,,
\end{equation}
i.e., near the black-hole horizon we select the {\it in-falling} solution. 
This latter condition is needed in order to obtain the retarded Green's function \cite{Son:2002sd,Policastro:2002se,Teaney:2006nc}, which is computed by differentiating twice the on-shell action~\eqref{action_gl} with respect to the source $\tilde X_0(\omega^2)$. 
The result is \cite{Colangelo:2009ra}
\begin{equation}\label{2pfunction}
\Pi^R_G(\omega^2)= \left. \frac{1}{k^\prime} \frac{f(u)}{z^3} e^{-\phi(z)} \tilde K(\omega^2,z) \partial_z \tilde K(\omega^2,z) \right|_{z=0} \,\,\, .
\end{equation}
Then, the spectral function is the imaginary part of the retarded Green's function
\begin{equation}\label{defSF}
 \rho(\omega^2)=\Im \left(\Pi^R_G(\omega^2)\right)\,.
\end{equation}
We have evaluated it for different values of temperature and chemical potential. 
Some numerical results are shown in Fig.~\ref{fig:SFglue}: in the left panel the first peak of the spectral function (multiplied by $k^\prime$) is plotted versus $\omega^2$ at $T=0.06c$ for different values of the chemical potential, whereas in the right panel the analogous curves computed at $\mu=0.109c$ for various values of temperature are plotted. 
As already observed at zero chemical potential in \cite{Colangelo:2009ra}, at any finite and fixed value of $\mu$ the peaks of the spectral functions broaden and move towards smaller values of $\omega^2$ as the temperature of the medium is increased, up to the melting point, where no peak can be distinguished anymore. 
The same behaviour is also observed at fixed temperature, increasing the chemical potential. 
The broadening is a sign that, as temperature and quark chemical potential increase, the states become unstable, getting a finite width.

\begin{figure}[h!]
 \centering
 \includegraphics[width=7cm]{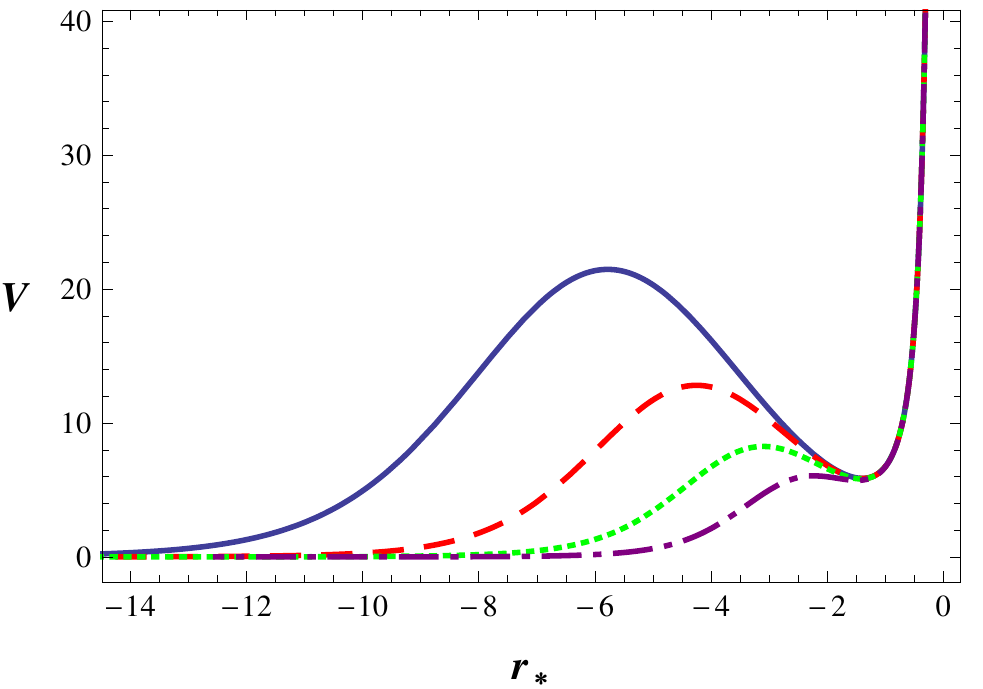}
 \caption{Potential in the Schr\"odinger equation of motion for the scalar glueball, Eq.(\ref{eqpsi}). The values of temperature and chemical potential are ($c=1$): $T=0.02$, $\mu=0.1$ (plain blue curve); $T=0.015$, $\mu=0.2$ (dashed red curve); $T=0.06$, $\mu=0.1$ (dotted green curve); $T=0.06$, $\mu=0.2$ (dash-dotted purple curve).}
 \label{fig:gluepot}
\end{figure}

This effect can also be appreciated by a quantum-mechanical argument.
Let us introduce the Regge-Wheeler tortoise coordinate $r_*$ \cite{Miranda:2009uw}, defined by the relation $\partial_{r_*}=-f(z)\partial_z$, $r_*\leqslant0$ ($r_*=0,-\infty$ correspond to $z=0,z_h$, respectively).
Then, through the Bogoliubov transformation $\tilde X=e^{B/2}\psi$, with $B(z)=z^2+3\log z$, the equation of motion for the glueball field $\tilde X$ becomes a Schr\"odinger equation for $\psi$:
\begin{equation}\label{eqpsi}
-\partial_{r_*}^2\psi(r_*)+V(r_*)\, \psi(r_*)=\omega^2\psi(r_*)
\end{equation}
with potential
\begin{equation}
V(z)=\left(\frac{B'(z)^2}{4}-\frac{B''(z)}{2}\right) f(z)^2-\frac{B'(z)}{2}\, f(z)\, f'(z)\,.
\end{equation}
The potential $V$ is plotted in Fig.~\ref{fig:gluepot} as a function of $r_*$. 
One can notice that at low $T$ and $\mu$, the potential has a well which makes the lowest eigenstates almost stable, while for higher values the decay probability becomes larger until the potential well disappears, as in \cite{Miranda:2009uw}.

\begin{figure}[h!]
 \centering
 \includegraphics[width=7cm]{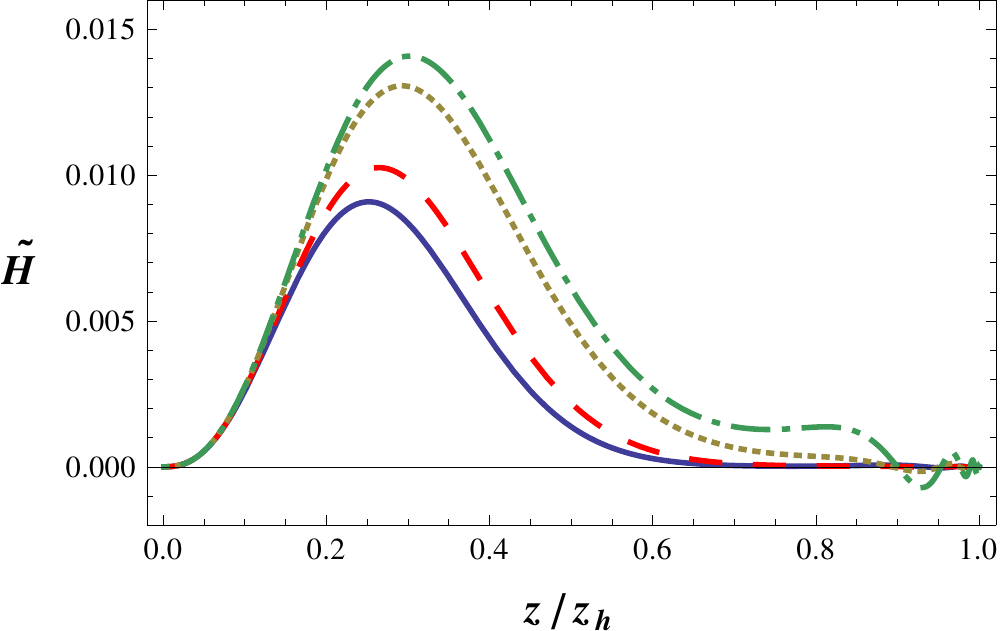}
 \caption{Glueball wavefunction at $T=0.05$ and low values of the chemical potential: $\mu=0.01$ (blue plain curve), $\mu=0.05$ (red dashed), $\mu=0.09$ (yellow dotted), $\mu=0.1$ (green dot-dashed, all quantities in units of $c$). The latter is unstable, therefore for $\mu\geqslant 0.1$ we cannot solve the eigenvalue problem to determine the glueball masses.}
 \label{fig:gluewf}
\end{figure}

In order to extract the dependence of glueball masses on the temperature and chemical potential, we have fitted each peak with a deformed Breit-Wigner function \cite{Colangelo:2009ra,Fujita:2009wc}:
\begin{equation}\label{BWeq}
\rho_{BW}(x)=\frac{a\,m\, \Gamma\, x^b}{(x-m^2)^2+m^2\Gamma^2}\,,
\end{equation}
where $m$ is the mass of the state and $\Gamma$ its width. 
We define the dissociation temperature as the lowest temperature at which the condition
\begin{equation}
\frac{h}{\Gamma^2}<1 \quad ,\quad \qquad h=\left.\frac{m\, \Gamma}{(x-m^2)^2+m^2\Gamma^2}\right|_{x=m^2}
\end{equation}
is fulfilled.
This means that we consider a state as melted if the ratio between the maximum of the pure Breit-Wigner function ($h$) and the squared width of the peak ($\Gamma^2$), the two parameters characterising the form of the peak in the $(m^2,\rho)$ plane, is lower than 1.
This criterion gives the same dissociation temperature at $\mu=0$ found in \cite{Colangelo:2009ra}, namely $T\sim 0.116c$.\\
At very low values of both $T$ and $\mu$, we compute masses as eigenvalues of the equation of motion, since in this case the black-hole is far away along the holographic coordinate axis ($z$) and it does not affect the wavefunctions in an appreciable way. 
We perform a different Bogoliubov transformation of the glueball field, $\tilde X(p,z)=e^{\hat B(z)/2}\tilde H(p,z)$, with $\hat B(z)=z^2+3\log z-\log f(z)$, such that the equation of motion for the field $\tilde H(p,z)$ reads:
\begin{eqnarray}\label{schreqglu}
&& -\partial_z^2\tilde H(\omega^2,z)+U(z)\,\tilde H(\omega^2,z)=\frac{\omega^2}{f(z)^2}\,\tilde H(\omega^2,z) \\
&& U(z)=\frac{\hat B'^2}{4}-\frac{\hat B''}{2}\,. 
\end{eqnarray}
The wavefunctions can be found by solving \eqref{schreqglu}, requiring that the eigenfunction vanishes at $z=0$ and for large $z$. 
In Fig.~\ref{fig:gluewf} some of them are plotted at fixed $T=0.05c$, for different values of the chemical potential.
When $\mu$ is larger than a certain value, the solutions cannot be considered as eigenfunctions anymore, since their oscillatory behaviour near the black-hole horizon becomes important.
We remark that the two methods described above to extract glueball masses give the same results for those values of $T$ and $\mu$ for which both can be applied.

\begin{figure}[h]
 \centering
 \subfigure{\includegraphics[width=6.7cm]{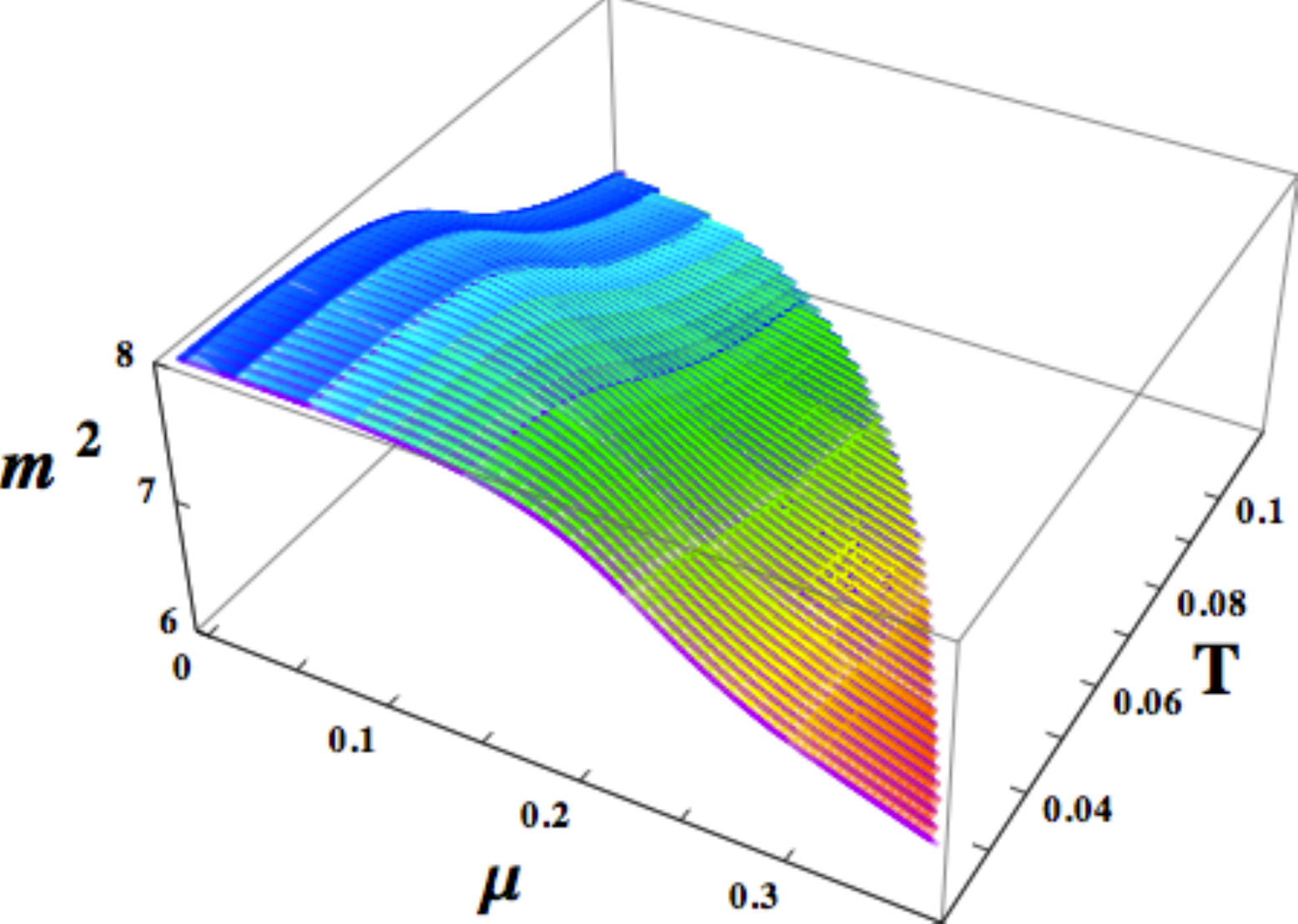}}
 \subfigure{\includegraphics[width=6.7cm]{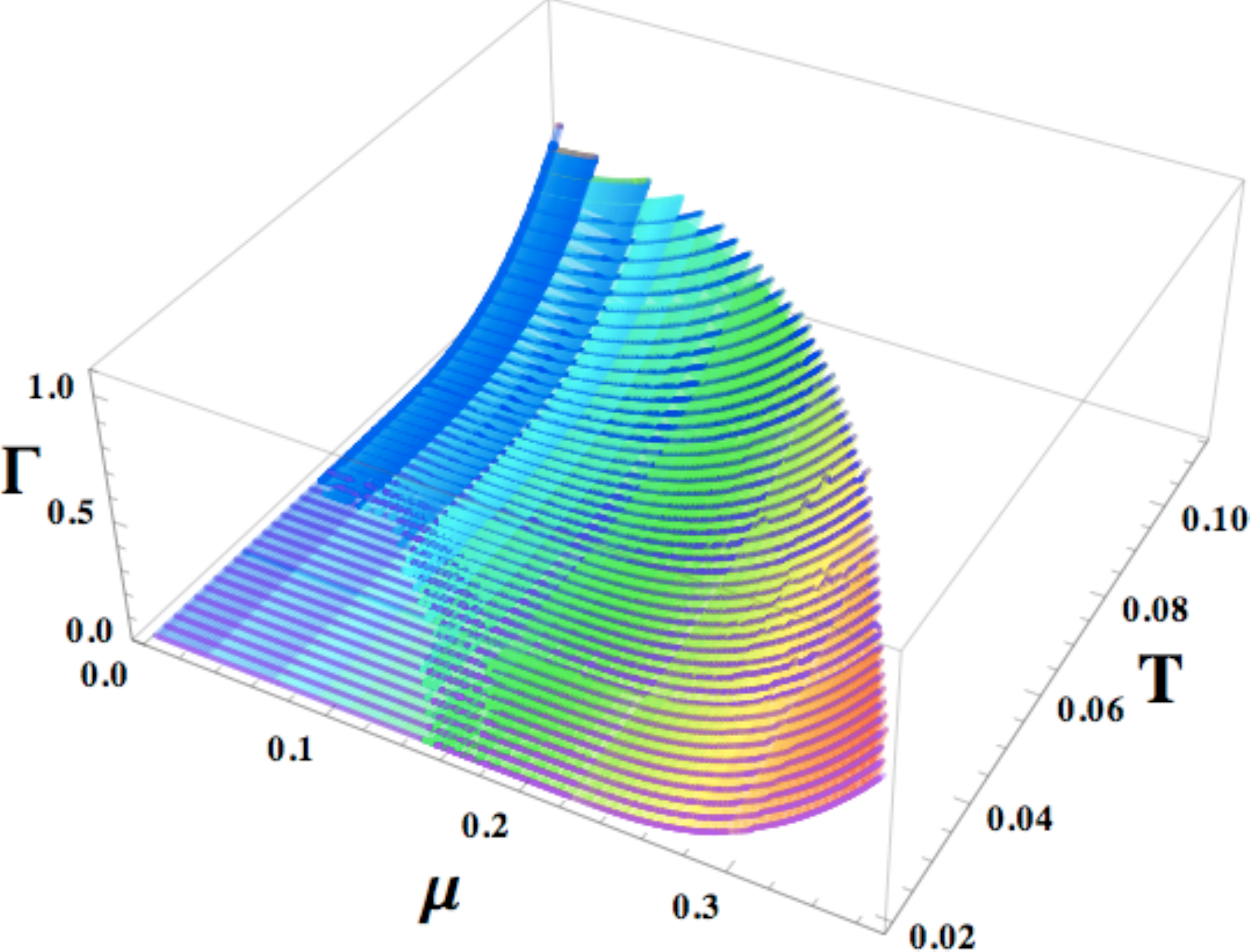}}
 \caption{Variation of the squared mass (left panel) and width (right panel) of the lightest scalar glueball with respect to temperature $T$ and chemical potential $\mu$. All quantities are in units of $c$.}
 \label{fig:gluemass}
\end{figure}

The variation of the squared mass and width of the lightest glueball is shown in Fig.~\ref{fig:gluemass}: we observe a decreasing of $m^2$ and an increasing of $\Gamma$ with temperature and chemical potential. 
This behaviour is similar to what has been obtained by an analysis of scalar glueball masses in a medium in $SU(2)_c$ lattice field theory \cite{Lombardo:2008vc}, in which a mass decrease at a very low temperature and high density in the transition from the hadronic to a superfluid phase has been predicted.

\section{Vector meson spectral function}\label{sec:vec}

An important application of the above described method is in the light vector meson sector. 
In fact, the $\rho$ meson is considered as a very useful probe to investigate modifications of the properties of the hot and dense phase, since it can be easily reconstructed and its decay products carry information about the medium in the laboratory \cite{Rapp:2010sj}.
 
\begin{figure}[b!]
 \centering
 \subfigure{\includegraphics[width=6.7cm]{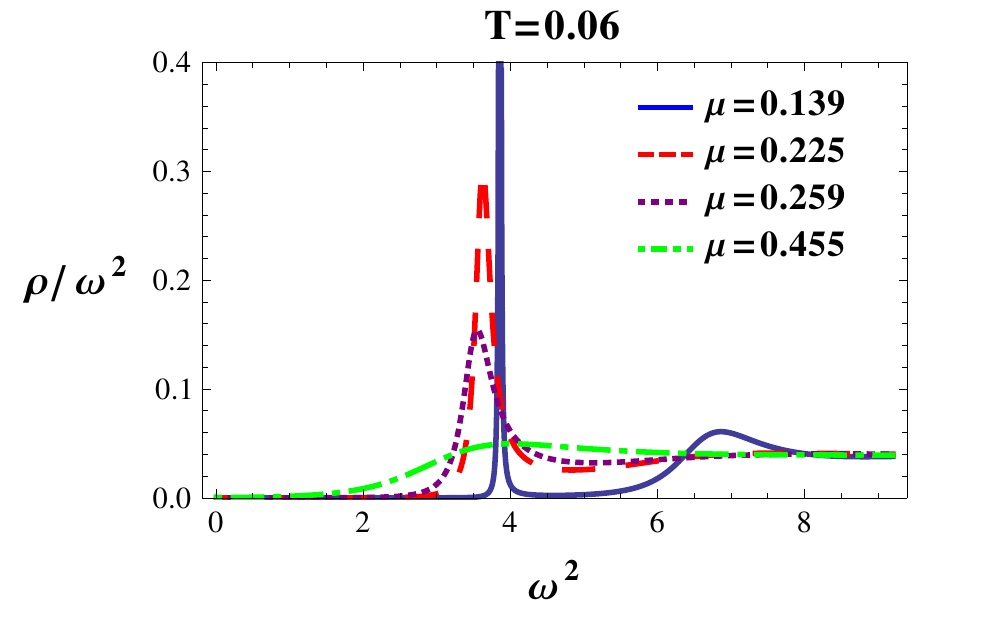}}
 \subfigure{\includegraphics[width=6.7cm]{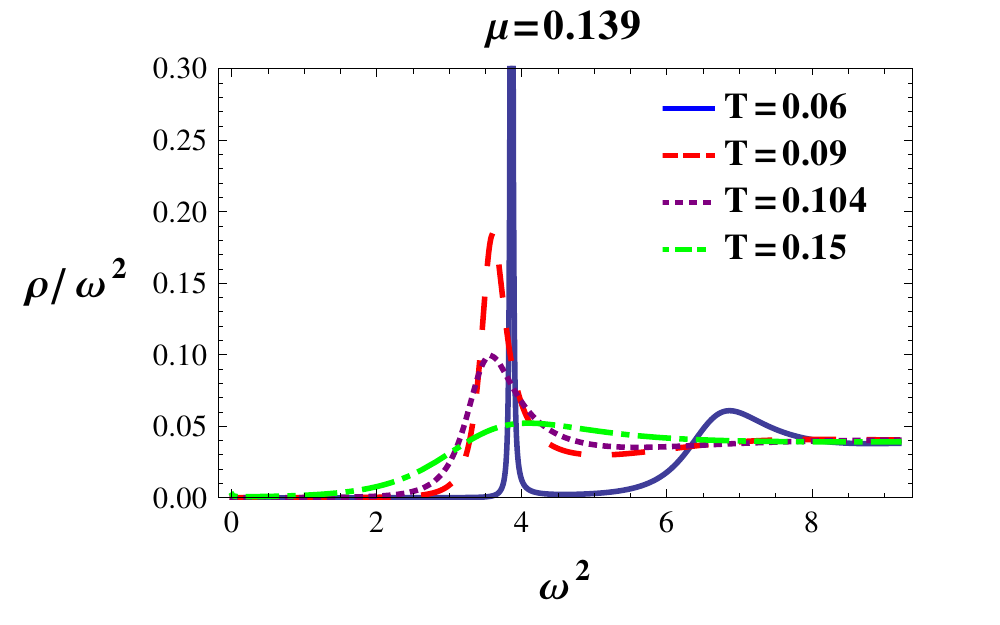}}
 \caption{Left: first two peaks of the vector meson spectral function at $T=0.06$ and $\mu=0.139$ (plain, blue), $\mu=0.225$ (dashed, red), $\mu=0.259$ (dotted, purple), $\mu=0.455$ (dot-dashed, green). Right: first two peaks of the spectral functions of vector mesons at $\mu=0.139$ and $T=0.06$ (plain, blue), $T=0.09$ (dashed, red), $T=0.104$ (dotted, purple), $T=0.15$ (dot-dashed, green). All quantities are in units of $c$.}
 \label{fig:SFvec}
\end{figure}

Following the soft-wall model formulation \cite{sw}, we introduce the gauge fields $A_{L}$ and $A_R$ associated to the $SU_L\otimes SU_R$ gauge group, dual to the left- and right-handed quark currents of QCD, the global 4$D$ symmetries being gauged in the bulk. 
The vector field, which is dual to the QCD operator $\bar q\gamma_\mu q$, is $V=(A_L+A_R)/2$, and the action for this field in the quadratic approximation is
\begin{equation}
 S=-\frac{1}{2\, k_V\, g_5^2}\int d^5x \sqrt{g}\, e^{-\phi(z)} \mbox{ Tr}\left[F_V^{MN}\, F_{V\, MN}\right]\,,
\end{equation}
where $F_V^{MN}=\partial^M\, V^N-\partial^N\, V^M$, with $V=V^aT^a$, $T^a$ being the generators of the $SU_V$ symmetry group; we shall choose the gauge $V_z=0$.
By comparing the two-point correlation functions of vector currents, computed in QCD and in the soft-wall model, one can fix $k_Vg_5^2=12\pi^2/N_c$ \cite{sw}.
The equation of motion for one of the three spatial components $V_i$ in the Fourier space is (setting $\bar p=0$)
\begin{equation}\label{eomvector}
 \partial_z\left( \frac{e^{-\phi(z)}}{z}f(z)\partial_z V_i(z,\omega^2)\right) + \frac{e^{-\phi(z)}}{z\, f(z)}\omega^2\, V_i(z,\omega^2)=0\,.
\end{equation}
We compute also in this case the retarded Green's function,
\begin{equation}
 G^R_{ij}(p_0)=\frac{\delta^2 S}{\delta V_i^0(-\omega)\delta V_j^0(\omega)}=\delta_{ij}\left. \frac{e^{-\phi(z)}\, f(z)}{g_5^2\, k_V} V(z,\omega^2)\frac{\partial_z\, V(z,\omega^2)}{z}\right|_{z=0}\,.
\end{equation}
 $V(z,\omega^2)$ is the bulk-to-boundary propagator of the vector field in the Fourier space, defined by $V_i(z,\omega^2)=V(z,\omega^2)\, V_i^0(\omega^2)$ and satisfying the equation of motion \eqref{eomvector}, with boundary conditions $V(0,\omega^2)=1$ and, as in the glueball case, the {\it in-falling} condition near the black-hole horizon:
\begin{equation}
V(z,\omega^2) \xrightarrow[]{z\to z_h} (1-z/z_h)^{-i\frac{\sqrt{\omega^2}\, z_h}{2(2-Q^2)}}\left(1+{\CMcal O}(1-z/z_h)\right)\,.
\end{equation}
Using Eq.~\eqref{defSF}, we find the spectral function in Fig.~\ref{fig:SFvec}, for some values of temperature and chemical potential; the first peak corresponds to the $\rho$ meson. 
In particular the quantity $\rho/\omega^2$ has been plotted since we expect the vector spectral funtion to increase quadratically for large values of $\omega$ (see, e.g. \cite{Ding:2010ga}).
Like scalar glueballs, also vector mesons become unstable and finally melt while temperature and quark chemical potential are increased. 
In fact, at high enough $T$ and/or $\mu$ the spectral function becomes flat, with no more resonances, reflecting the formation of a weakly interacting plasma of quarks and gluons.
Looking at Fig.~\ref{fig:SFvec}, one can notice that the dissociation of the first excited state occurs much earlier than the one of the ground state (the $\rho$ meson).

\begin{figure}[h!]
 \centering
\subfigure{\includegraphics[width=6.7cm]{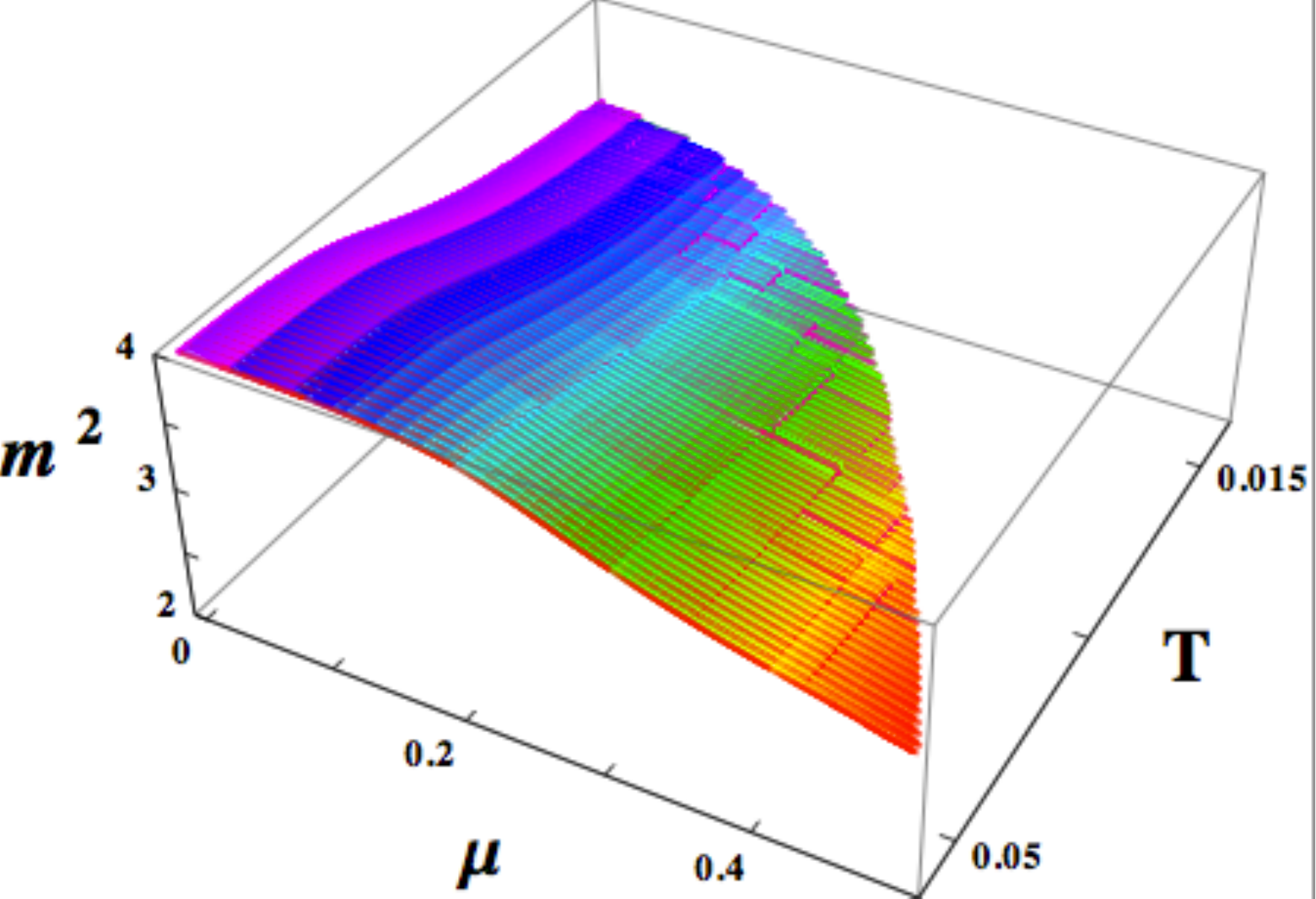}}
\subfigure{\includegraphics[width=6.7cm]{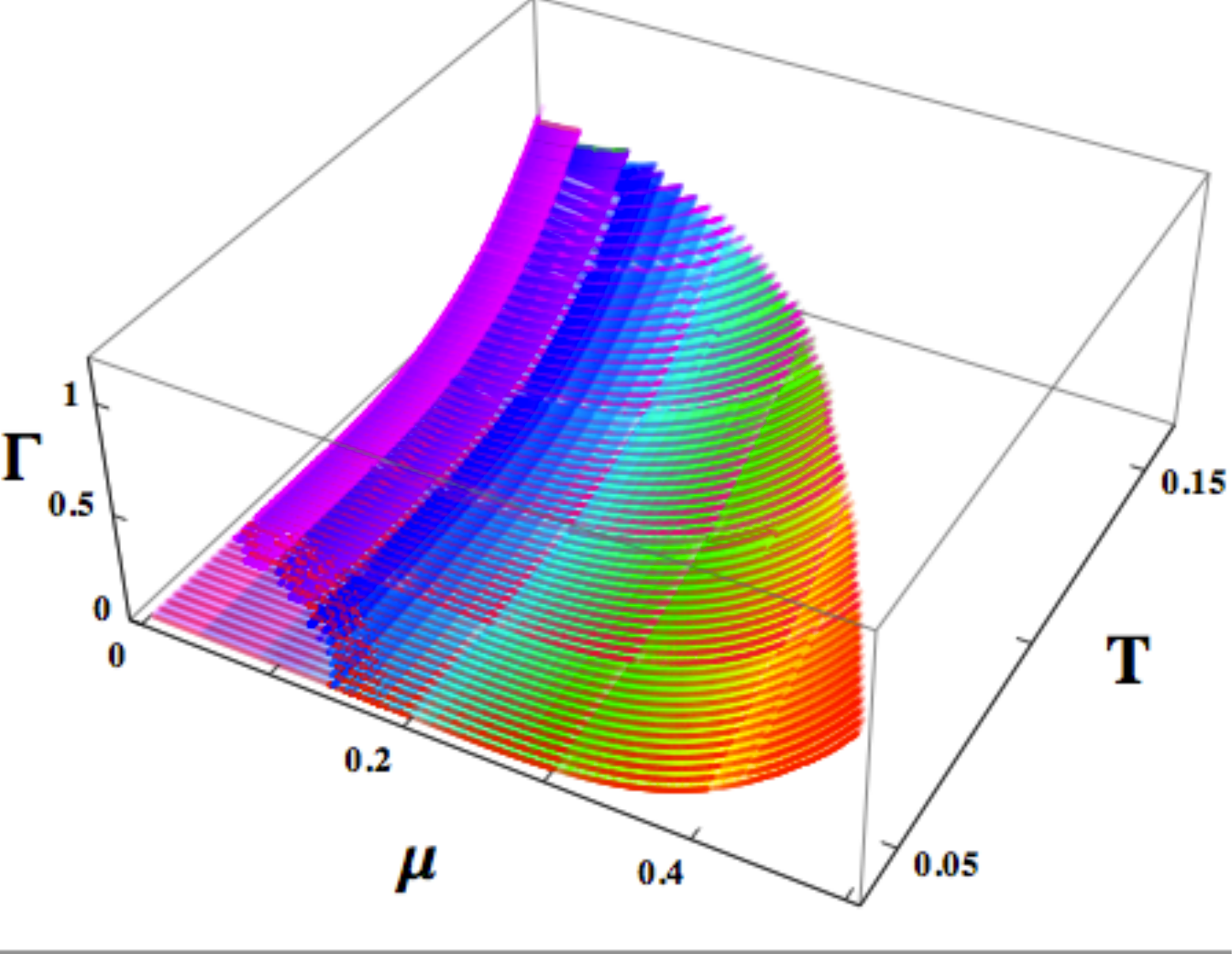}}
 \caption{Variation of the squared mass (left panel) and width (right panel) of $\rho$ meson versus temperature $T$ and chemical potential $\mu$. All quantities are in units of $c$.}
 \label{fig:vecmass}
\end{figure}

The same analysis carried out in the glueball sector can be applied to vectors: using Eq.~\eqref{BWeq} it is possible to calculate the position of the peaks (masses) and their width, which are shown in Fig.~\ref{fig:vecmass}. 
Again, at low values of the two parameters $\mu$ and $T$, the widths vanish and masses are computed by solving the equation of motion for the vector fields in the Schr\"odinger form.
\\
Fig.~\ref{fig:vecmass} shows that the $\rho$ mass decreases with increasing temperature and chemical potential. 
In particular, at $T\sim0$ we observe a $13\%$ effect ($25\%$ for the squared mass), while at an intermediate temperature $T=0.1c$ the effect reduces to $8\%$ ($16\%$ for the squared mass).

In the chiral model studied in \cite{Brown:1991kk} to describe the in-medium behaviour of mesons, the scaling of the $\rho$-meson mass has been related to parameters governing chiral symmetry breaking.
In particular, along with the dropping of the chiral condensate at finite density and temperature, a $\sim20\%$ decrease of the $\rho$ mass has been found at nuclear matter densities.
In the soft-wall model both vector meson masses and the chiral condensate \cite{Colangelo:2011sr} decrease at increasing temperature and density, qualitatively in agreement with the Brown-Rho picture, suggesting the existence of a relation between these two phenomena.
In our holographic model, chiral symmetry breaking and vector meson masses are two distinct sectors and a direct relationship probing such a scaling cannot be explicitly found:
nevertheless, a comparison between the finite density behaviour of the pion decay constant $f_\pi$ and the $\rho$ meson mass can be carried out, but deserves a separate study.

In \cite{Hatsuda:1991ez}, Hatsuda and Lee use QCD sum rules to obtain the density dependence of the $\rho$-meson mass; they find a decreasing behaviour fitted by a linear function: $m(\rho)/m(0)=1-\alpha\,\rho/\rho_0$, where $\rho$ is the density ($\rho_0=0.16$~fm$^{-3}$ is the nuclear matter density), and their prediction for $\alpha$ is $\alpha=0.16\pm0.06$.
Also in \cite{Yokokawa:2002pw} it is suggested that in-medium hadronic spectral functions carry important information on chiral structure of hot/dense matter. 
By studying unitarized chiral models, a softening of $\sigma$ and $\rho$ mesons is found, which is driven by the decrease of $f_\pi$ in nuclear matter.
In \cite{Bertin:1990ec} a quark model of nuclear matter is put forward, in which quarks interact with the nuclear field.
As a result, a downward shift of $120$ MeV has been predicted for the $\rho$ mass. 
A mass dropping of $\rho$ mesons has also been found in \cite{Post:2003hu} within a coupled-channel analysis, in \cite{Santini:2008pk} within an extended vector meson dominance model, in \cite{DuttMazumder:2000ys} within sum rules at finite nucleon density, and in lattice SU(2) calculations \cite{Muroya:2002ry,Hands:2007uc}.
However the behaviour of hadrons in medium is still a debated issue, both experimentally and theoretically. 
While the broadening of the spectral functions is widely accepted, the variation of the masses is unclear. 
Some data agree with a small decrease \cite{Huber:2003pu,Naruki:2005kd,Adamova:2006nu}, and other show a very small effect, even compatible with zero \cite{Arnaldi:2006jq,Wood:2008ee}.
In particular, in \cite{Huber:2003pu} an analysis of the experimental data through various theoretical models has been carried out.
In \cite{Naruki:2005kd}, a $\sim9\%$ decrease in mass for the $\rho$ meson at normal nuclear density is found from a linear fit, as suggested by the Hatsuda-Lee model \cite{Hatsuda:1991ez} ($\alpha=0.092\pm0.002$).
In \cite{Arnaldi:2006jq} a modification of the $\rho$ spectral function is reported which is not compatible with a mass shift, but only with a broadening of the corresponding peak.
In the analysis performed by the CLAS Collaboration \cite{Wood:2008ee} both the mass shifts by the Brown-Rho \cite{Brown:1991kk} and the Hatsuda-Lee \cite{Hatsuda:1991ez} models are ruled out, since no evidence of large modifications to the $\rho$ mass has been found ($\alpha=0.02\pm0.02$), and an upper limit for $\alpha$ has been set to $\alpha=0.053$ with a 95$\%$ confidence level. 

The prediction of this model for the shift parameter $\alpha$ is $\alpha=(m(0)-m(\mu_0))/m(0)\simeq0.012$ at $T=0.023c=9$~MeV, where $\mu_0=0.209c=81.1$~MeV is the value of the quark chemical potential corresponding to the nuclear matter density $\rho_0$.
However, in the present model the quark density $\rho=\kappa q$ is related to the chemical potential in a model dependent way (through $\kappa$), and again we have used $\kappa=1$.
Although at values of $\mu$ corresponding to the dissociation of the $\rho$ meson we find a $\sim13\%$ decrease of its mass, for $\mu$ corresponding to nuclear matter density the effect is only $\sim1\%$, in agreement with the experimental analyses that find a small or compatible with zero effect.

There are also theoretical studies predicting an increase in mass \cite{Urban:1999im,Cabrera:2000dx}. 
A similar result has been found in the hard-wall model, another bottom-up holographic model characterised by a different mechanism of conformal symmetry breaking \cite{Erlich:2005qh}; in particular, an increase of meson masses with density at $T=0$ is found by solving the corresponding equations of motion \cite{Jo:2009xr}.
This behaviour, which is in contrast with the one found in our analysis and with the experimental results described above, is similar to the one found in other top-down holographic approaches \cite{Mas:2008jz}.

Finally, we remark that it would be interesting to extend the calculations based on the holographic approach to heavy mesons like $J/\psi$, investigating the behaviour of these states in a hot and dense medium; some methods have been proposed about how to describe charmonium-like states in the soft-wall model \cite{Fujita:2009wc,Grigoryan:2010pj}.

\section{Conclusions}

Studying scalar glueballs and light vector mesons in a hot and dense medium, we have observed a broadening of the peaks in their spectral function, which is a signal that the states become unstable as temperature and chemical potential are increased, and a shift of the position of the peaks towards smaller values of the mass, signaling a decrease of the hadron masses. 
While the first effect has also been found in other theoretical studies and in some experimental data, the latter is still a debated issue. 
We have also found that at $T=0$ the melting of glueballs and vectors occurs at $\mu\sim0.365\kappa c$ and $\mu\sim0.499 \kappa c$, respectively. 
If we set $\kappa=1/2$, as in \cite{Colangelo:2010pe}, the numerical values of the chemical potential are $\mu\simeq71$~MeV for the glueball and $\mu\simeq97$~MeV for the vector mesons.
At increasing temperature, these critical values decrease, and at $T\sim0.114c$ ($T\sim44$~MeV) and $T\sim0.162c$ ($T\sim63$~MeV) the spectral function of glueballs and vectors, respectively, is almost flat already at zero density.
Applying the soft-wall model, inspired by the AdS/QCD correspondence, we can address the nonperturbative problem with little effort, numerically solving a linear differential equation, and trying to collect as much information as such a phenomenological model of holographic QCD can provide.
The broadening of the peaks confirms that the model captures an important accepted feature of QCD.
The decreasing of the masses is a prediction which, particularly in the case of vector mesons, can be tested in the experiments;  in the case of the density dependence, such a prediction,  obtained in the soft-wall model,  is different from the one obtained in other holographic approaches.

\vspace*{1cm}
\noindent{\bf Acknowledgments.} We thank F. De Fazio for useful discussions.
This work is supported in part by the Italian MIUR PRIN 2009.

\end{document}